# Predicting the Realizable Maximum Power Factor using the *Jonker* and *Ioffe* formulation: Al-doped ZnO Triangular Microcrystals with Graphite Inclusion Case Study

Soumya Biswas[1], Keshav Dabral[2], Saptak Majumder[1], Rajasekar Parasuraman[3], Aditya S. Dutt[4], Vinayak B. Kamble[1,*]

[1]School of Physics, Indian Institute of Science Education and Research Thiruvananthapuram India 695551.

[2]Department of Physics, Graphic Era Hill University, Dehradun, Uttarakhand, India 248002

[3]Department of Chemistry, School of Advanced Sciences, Vellore Institute of Technology (VIT), Vellore, Tamil Nadu 632 014, India.

[4]Interuniversity Microelectronics Centre, Kapeldreef 75, 3001 Leuven, Belgium.

**\*Email: kbvinayak@iisertvm.ac.in**

Supporting information:

# Predicting the realizable maximum power factor using the *Jonker* and *Ioffe* formulation: Al-doped ZnO Triangular microcrystals with graphite inclusion case study

Soumya Biswas[1], Keshav Dabral[2], Saptak Majumder[1], Rajasekar Parasuraman[3], Aditya S. Dutt[4], Vinayak B. Kamble[1,*]

[1]School of Physics, Indian Institute of Science Education and Research Thiruvananthapuram India 695551.

[2]Department of Physics, Graphic Era Hill University, Dehradun, Uttarakhand, India 248002

[3]Department of Chemistry, School of Advanced Sciences, VIT University, Vellore, India 632014

[4]Interuniversity Microelectronics Centre, Kapeldreef 75, 3001 Leuven, Belgium.

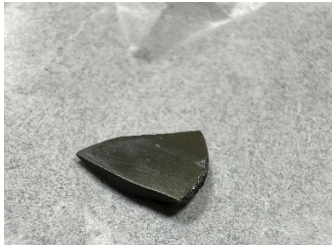

**Figure S1.** A digital picture of cross section of a cut-piece from ZnOGr pellet.

## 1. Further analysis of XRD data

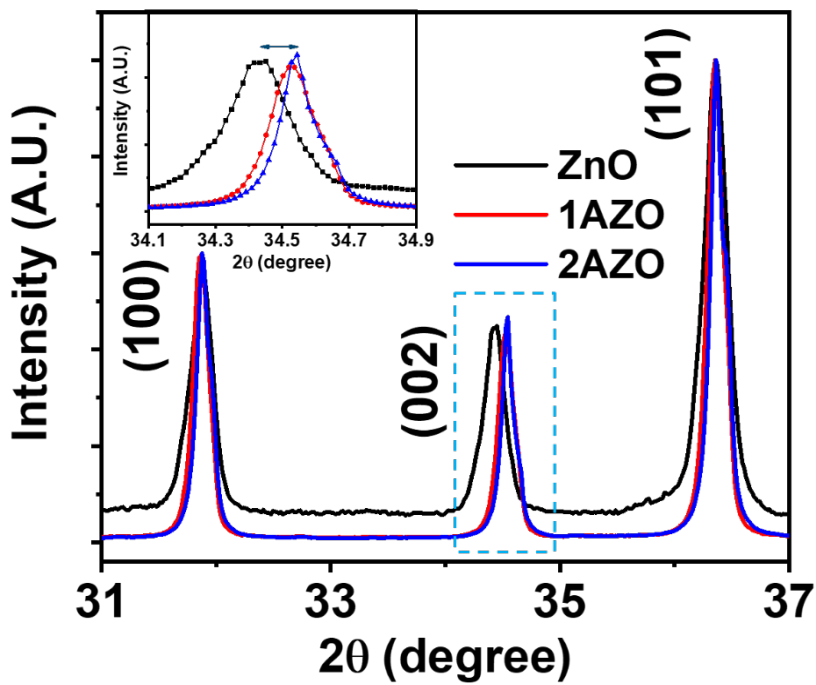

**Figure. S2** The magnified XRD pattern of doped and undoped ZnO samples without graphite to confirm th doping. In the inset, the shift is shown for the (002) peak.

The figure S1 shows the comparison of the the XRD plots of the three ZnO samples (undoped, 1% and 2% Al doped). The inset shows the enlarged view of the (002) line that shows maximul change. It may be noted here that the undoped ZnO peak has the highest and symmetric broadening on the other hand the doped ones show smaller and asymmetric broading The peak asymmetry may arise from the doping induced compressive strain in the lattice given the ionic radii mismatch of the two cations.

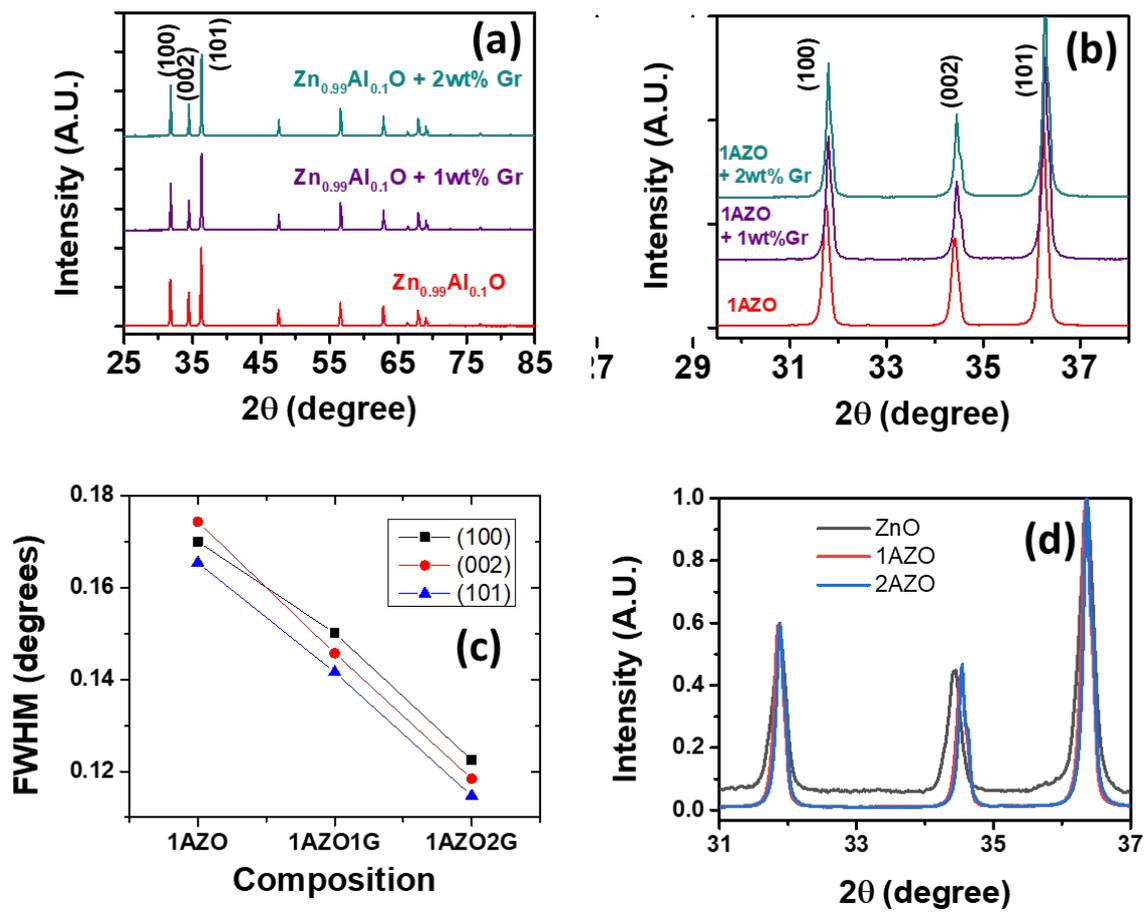

**Figure. S3**. The XRD pattern of (a) 1% Al-doped ZnO with different graphite concentrations and (b) the magnified (100) XRD peak of the same samples. (c) Table of the FWHM of 3 main peaks of the same samples.

## 2. Microstructure of the Al doped ZnO with Graphite inclusions

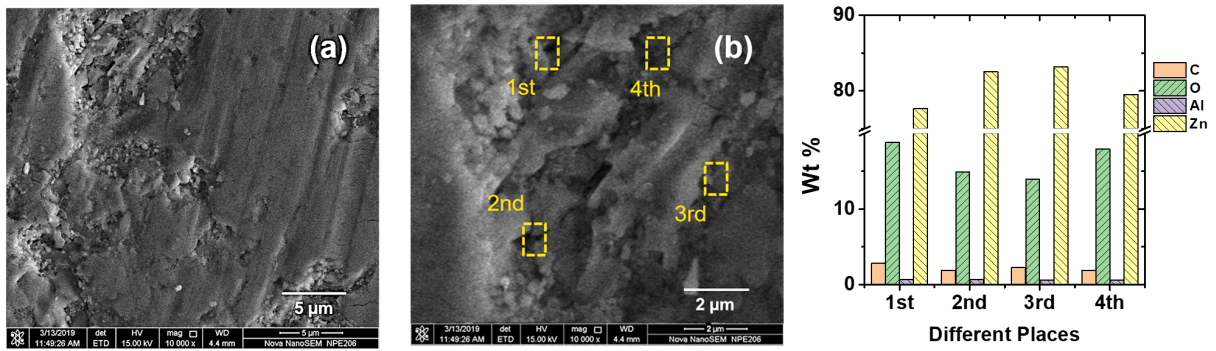

**Figure. S4**: The (a) SEM micrograph and (b) magnified SEM micrograph of the 1AZO1G broken sample. (At the marked positions the SEM-EDS is performed) (c) The wt % of compositions in different marked positions.

The samples were broken from the middle and SEM as well as EDS is performed on the interior side of the 1% Al-doped ZnO having 2 wt% Gr composition. The SEM micrographs are shown in Fig. S3(a) and (b). The SEM-EDS is performed at the different places of the 1AZO2G sample as shown in Fig. S3(a) and (b). It is performed mainly to confirm the existence of carbon. The Wt% of the compositions is shown in the Fig. S3(c) that reflects presence of carbon and almost uniform throughout the sample. This confirms the homogenous nature of the samples with existence of graphite.

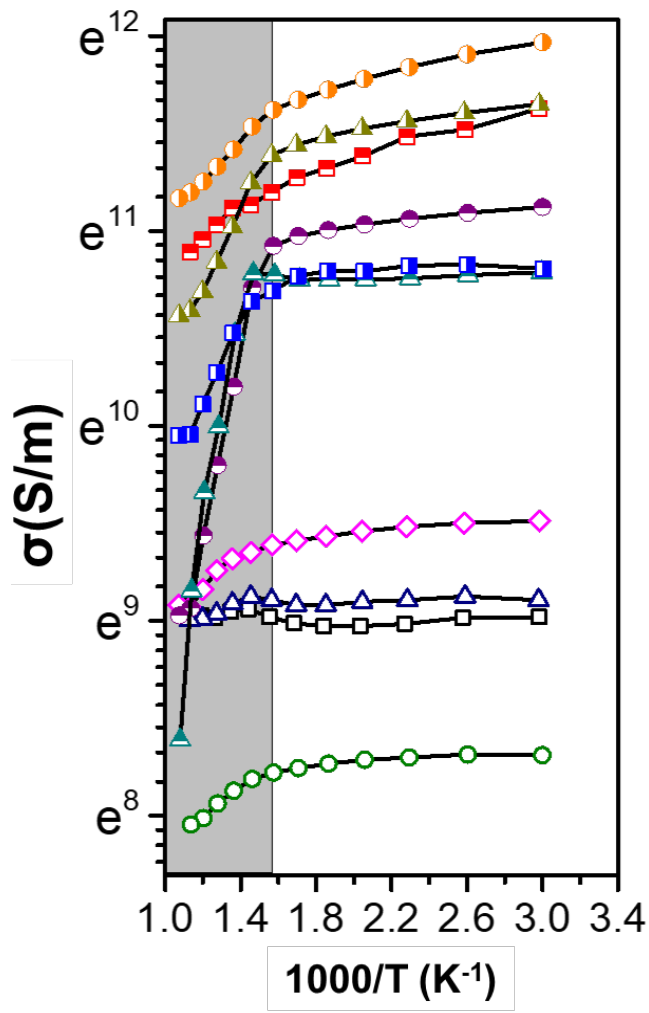

**Figure S5**. The variation in conductivity in the natural log scale with 1000/temperature for all the ZnO samples with different concentrations of Al doping and graphite addition.

**Table S1. Sample code and activation energy of the prepared samples in two temperature ranges.**

| Sample Specification | Sample Code | Activation Energy at high T (625 K - 1000 K) meV | Activation Energy at low T (300 K - 625 K) meV |
|---|---|---|---|
| Bare ZnO | ZnO | 2.12 | 2.12 |
| ZnO + 0.5wt% graphite | ZnO0.5G | -68.8 | -8.4 |
| ZnO + 1 wt% graphite | ZnO1G | -62.3 | -6.3 |
| ZnO + 2 wt% graphite | ZnO2G | -33.3 | -0.6 |
| 1% Al doped ZnO | 1AZO | -65.8 | -10.9 |
| 1%Al-doped ZnO + 1 wt% graphite | 1AZO1G | -380.5 | -16.8 |
| 1%Al-doped ZnO + 2 wt% graphite | 1AZO2G | -490.6 | -0.8 |
| 2% Al doped ZnO | 2AZO | -161.5 | -7.9 |
| 2% Al-doped ZnO + 1 wt% graphite | 2AZO1G | -80.7 | -22.3 |
| 2% Al-doped ZnO + 2 wt% graphite | 2AZO2G | -154 | -18.6 |

For intrinsic semiconductors, we can write the Fermi level by the formula:

$$E_{Fi} = \frac{E_c + E_v}{2} + \frac{kT}{2} \ln\left(\frac{N_v}{N_c}\right)$$

Where $E_c$ is the energy of the conduction band, $E_v$ is the energy of the valance band, k is the Boltzmann constant, T is absolute temperature, $N_c$ is the conduction band density of states and $N_v$ is the valance band density of states.

For n-type extrinsic semiconductors, we can write the Fermi level by the formula:

$$E_F = E_{Fi} + kT\ln\left(\frac{N_D}{n_i}\right)$$

Where $N_D$ is the total number of donors and $n_i$ is the intrinsic carrier concentration.

By combining these two equations we can write the Fermi-level position by the following equation:

$$E_F = \frac{E_c + E_v}{2} + \frac{kT}{2} \ln\left(\frac{N_v}{N_c}\right) + kT\ln\left(\frac{N_D}{n_i}\right)$$

Now taking into account the $E_c = 0$ will provide the relative position of the fermi level from the conduction band. At T=300K, kT= 25 meV, $E_v$ = -3.4 eV. The $N_v$ and $N_c$ for ZnO are respectively of the order of $10^{19}$ per cm³ and 2.2 x $10^{18}$ cm³. The $N_D$ is $10^{19}$ per cm³ and $n_i$ is of the order of $10^5$. Putting these quantities in the equation the position of the Fermi level relative to the conduction band can be calculated approximately. The position is found to be around 860 meV below the conduction band.

When it is calculated taking $N_D$ around $10^{20}$ the position is found to be around 800 meV below the conduction band.

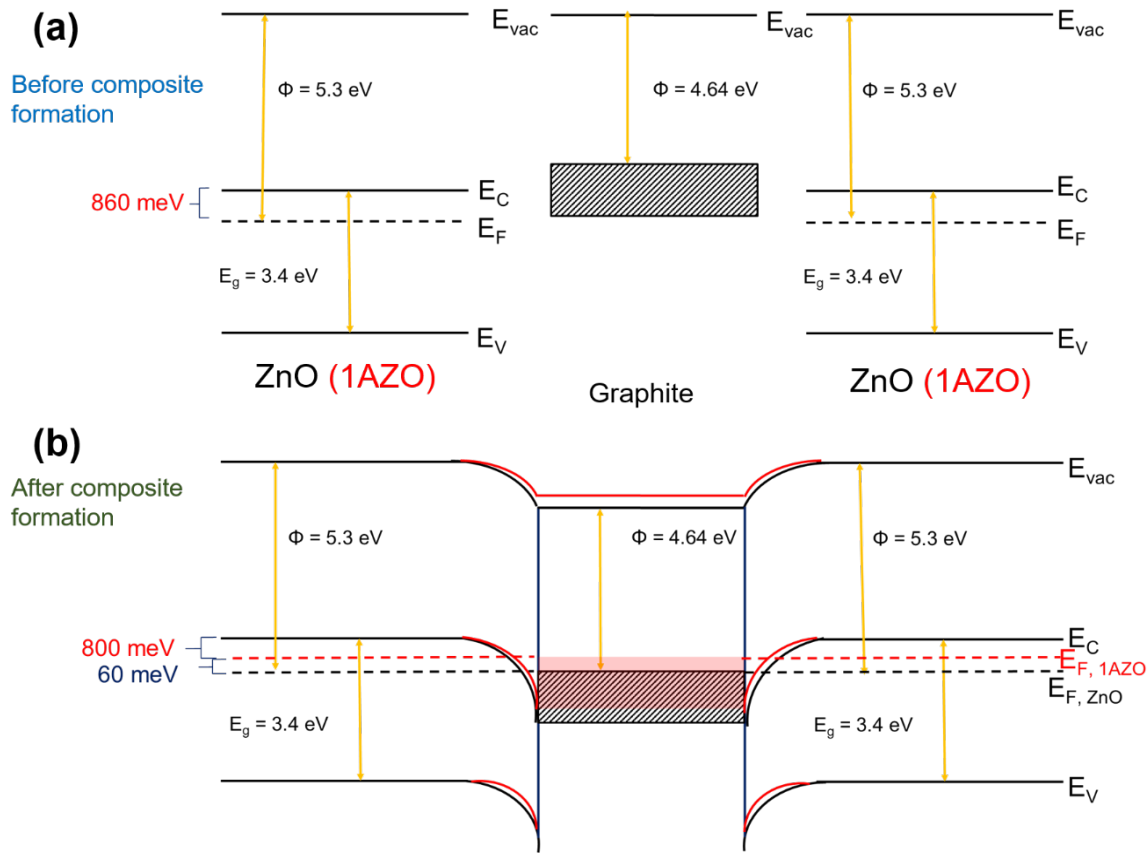

**Figure S6.** Band diagram of ZnO-graphite (a) before and (b) after the composite formation. (The red lines correspond to the 1AZO sample with graphite).

The band bending of the undoped and 1% doped ZnO with graphite is shown in Fig. S5. Fig. S5 (a) shows the band diagram before the formation of the composite and Fig. S5 (b) shows the band diagram after the composite formation. The red lines correspond to the 1% Al-doped ZnO with graphite samples. The calculation of the Fermi level confirms the band bending explained in the electrical conductivity section.

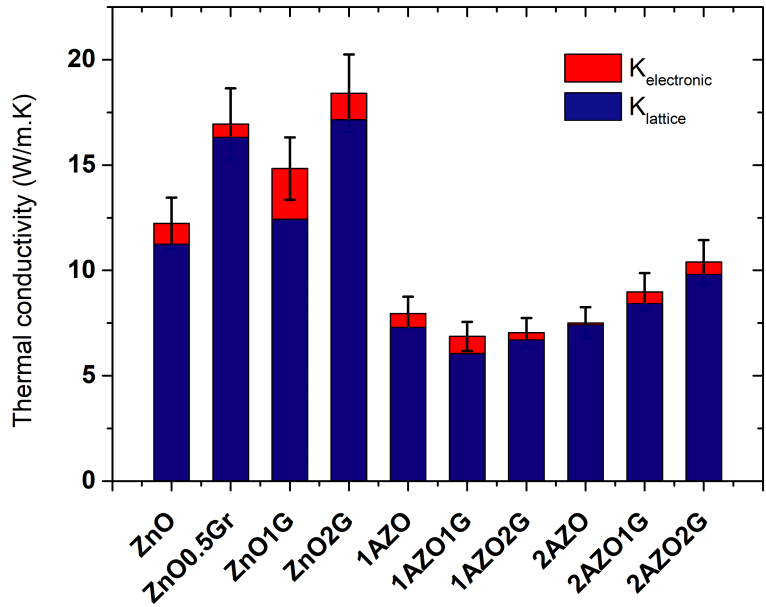

**Figure S7.** The comparison of the near room temperature (335 K) thermal conductivity of the samples.

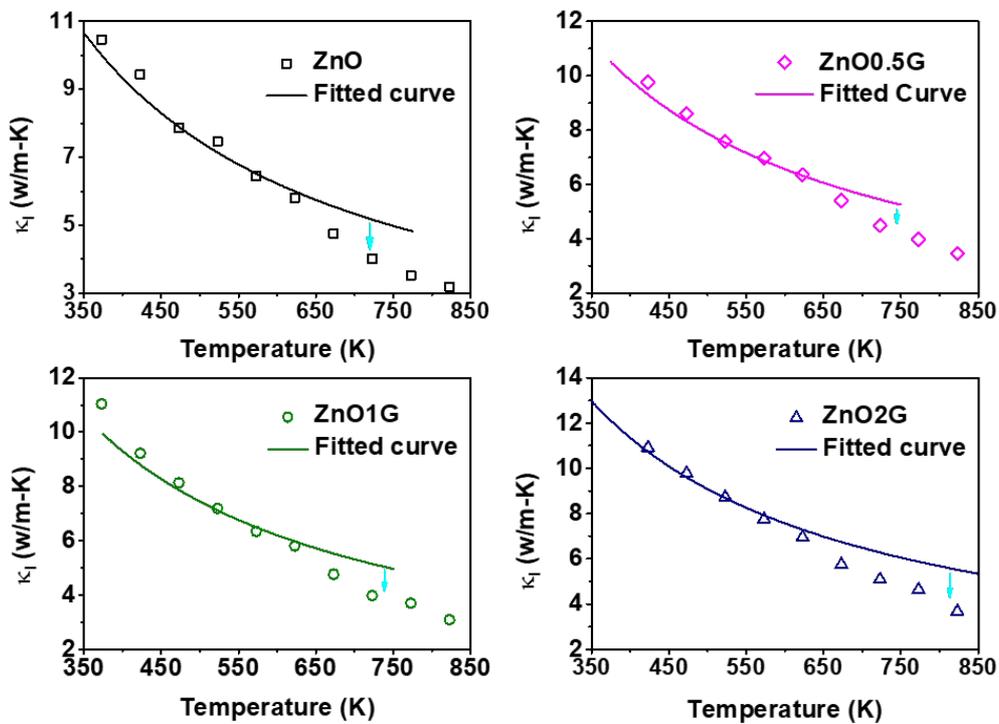

**Figure S8.** The measured and fitted lattice thermal conductivity data of the undoped ZnO samples with graphite mixture.

**Abstract**

Among the popular TE materials, selenides and tellurides are the benchmarks of high-efficiency systems. However, for the high-temperature application (>700 K), it is required to rely on the silicides and the oxides due to their exceptional thermal stability. ZnO is among the first few oxides in the field of thermoelectricity. Al-doped ZnO is a proven material for its high-temperature thermoelectric applications. However, the high grain boundary resistance limits further improvement of the efficiency of this oxide. Band-engineering, band-modification is a successful approach in lowering the grain boundary resistance. The addition of graphite and graphite-based materials at the grain boundaries is shown to serve this purpose. In this work, graphite powder is added in varying proportions to Al-doped ZnO triangular microcrystals. Thus, prepared materials are characterized to confirm the formation and investigate the nature of interface, morphology, etc. TE parameters such as electrical conductivity, Seebeck coefficient, and thermal conductivity of those materials also have been measured. The theoretical calculation of TE efficiency zT often differs from the actual experimental results due to the wide range of preparation methods, leading to changes in porosity, the nature and density defects, and several other factors. In this paper, an effort has been made to estimate the maximum achievable power factor ($PF_{max}$) from the measured TE parameters of this set of samples by the *Jonker* and *Ioffe* analysis. Based on the predicted $PF_{max}$, an appropriate material composition has been identified to achieve that same. Subsequently, including the measured parameters the TE efficiency (zT) is calculated. Further, a sudden dip observed in the thermal conductivity at the high-temperature range (625 K - 1000 K) of the prepared undoped ZnO graphite composite is investigated in this paper.

**Keywords**

## 1. Introduction

In the present scenario of excessively relying on perishable fossil fuels, it is imperative to search for new renewable energy sources. One of the safest and easiest economic processes is to convert the heat that otherwise goes to waste, into electricity by the use of thermoelectric (TE) power generation.[1,2] There are several occasions where a significant amount of heat is generated, dissipated in nature, and barely recovered. With the help of a thermoelectric generator, it is possible to transform this waste heat into electrical energy,[3,4] which in terms will help to solve the problem of renewable energy resources. For these TE generators, an efficient TE material is also required. The material should not only be efficient but also must be environment-friendly, affordable, and chemically stable.

There are a few efficient materials like GeTe,[5] $Bi_2Te_3$,[6–8] PbTe,[9] SnSe,[10,11] etc. some of which are toxic or less earth-abundant and unstable at high temperatures. This has created the need to search for more earth-abundant, non-toxic, as well as high-temperature stable thermoelectric materials. Due to these reasons, the scientific community is on a quest to discover efficient materials in oxides,[12] silicides,[13] etc.; which are chemically stable and easy to synthesize. There are several efficient thermoelectric oxides such as ZnO,[14–17] $SrTiO_3$,[18] $In_2O_3$,[19,20] complex cobaltites,[21–23] etc. However, these materials are barely conducting and the thermal conductivity is moderate to high. The main challenge remains to increase the efficiency of these materials by microstructure modification. Several methodologies have been used to modify these materials suitably which include doping with donor or acceptor impurities[24] to boost the carrier density, nano-structurization,[25] introductions of defects,[2,6] etc. to lower the thermal conductivity utilizing enhanced phonon scattering, energy, and carrier filtering,[27] band engineering,[28] spin state modification,[29] etc. In some instances, these successfully have achieved the goal. Enhancing the power factor (PF) is one way to increase efficiency. The formation of composite structures has been proven to be one of the most successful strategies for improving the TE properties of ZnO.[15,30,31] In nanocomposites, the manipulation of electronic transport can be achieved by engineering interfaces between the different materials that lead to various processes like energy filtering, charge carrier mobility optimization, and phonon scattering contributions. From a thermoelectric perspective, the formation of composites most often optimizes the power factor ($S^2\sigma$) and lowers thermal conductivity. The lowering of thermal conductivity arises from enhanced grain-boundary scattering, impurity or inclusion scattering, and lattice softening. Mostly in oxides and a few chalcogenides, the presence of surface defect states introduces the grain boundary resistance that could be avoided by suitable surface capping with other materials like Graphene or its derivatives. The introduction of a suitable material at the grain boundaries assists in lowering the grain boundary resistance thereby achieving band engineering in materials like ZnO,[17] $SrTiO_3$,[32] SnSe[33] etc. Carbon-based materials such as Graphene, Reduced Graphene oxide, graphite, Carbon Nanotubes, etc. have been explored for the formation of composite with chalcogenides as well as oxides and enhance their TE performances.[15,34,35] Recent research on Nb-doped $SrTiO_3$ with graphite has demonstrated enhanced electrical conductivity by an order of magnitude.[36] Similarly, a 4-fold improvement in the TE figure of merit (zT) has been reported when graphite has been used to form a composite with another popular TE oxide $TiO_2$.[37] In the case of the oxide materials high grain boundary resistance has been identified as one of the biggest reasons for their poor electrical conductivity.[38] The introduction of a suitable work function material between the two ZnO grains such as RGO, graphite, or metal with a lower work function than ZnO can bend the bands of ZnO downward at the grain boundaries allowing more carriers to cross the barrier. Thus, it reduces the grain boundary resistance. However, the extent of reduction of grain boundary resistance is governed by the nature of the composite. The overall conductivity of the composite material depends on the amount of the additive material used, the change in carrier concentration, mobility, temperatures, etc. However, the inclusion of graphite not only led to a reduction of the grain boundary resistance but also formed a few parallel high-conductive conductive graphite paths to smoothen the movement of the electrons without affecting the Seebeck coefficient substantially. As a result, an overall TE performance enhancement has been observed.[15,37,39] Chen et al.[15] and several others have already reported a significant carrier concentration rise in the Al-doped ZnO-RGO composite, thereby improving TE performance. However, the effect of more conductive Graphite has been envisaged to study their effect on the TE transport properties of ZnO-graphite composite. In the case of the modification of one material by introducing one or more others as a dopant or as an additive, it is challenging to optimize these dopants/additives concentrations to improve the PF. Theoretical calculations may predict the amount of these additives up to a certain extent but in the actual experiment, it hardly meets the expectations due to processing conditions affecting the microstructure strongly. Therefore, it is required to find a coherent way to bring about this modification. To start with, several materials with the addition of foreign materials can be prepared with different concentrations. The thermoelectric properties, mainly the Seebeck coefficients S and the electrical conductivity values $\sigma$ are to be measured. Subsequently, the *Jonker* and *Ioffe* analysis[40] may be performed based on those measured data and predict the optimized PF of those sets of materials.

ZnO is one of the strong contenders among these oxides-based thermoelectrics. It has been widely investigated for decades.[14,16,41,42] The wide band gap,[43] adverse phonon dispersion[44], and the electronic structure of ZnO support high thermal conductivity and less electrical conductivity. Though reducing the thermal conductivity is helpful, it is more convenient to optimize the power factor, which is a combination of the electrical conductivity and the Seebeck coefficient. In the previous work[45], an unintentional addition of graphite during the hot-pressing stage has given a high-power factor. As a continuation of that work, the effort has been made to mix the graphite with the ZnO grains intentionally to optimize the power factor as well as the efficiency of the ZnO and to investigate the effect of mixing graphite with the ZnO. In this work, the first few samples of different Al doping concentrations in ZnO triangular microcrystal powder have been synthesized. Thereafter graphite of different wt% has been mixed with those doped and undoped ZnO powders. Later on, using the *Jonker* and *Ioffe* analysis, an attempt is made to predict the highest achievable power factor of the samples. Based on the trend of the initial experiments,[45] samples of different graphite concentrations with undoped ZnO powder have been made and coincidentally it has reached the well-desired PF as well as the highest efficiency among those synthesized materials. Therefore, in this methodology, an idea has been conjectured to understand the highest possible PF value of unstudied materials experimentally. On the other hand, beginning with new material mixtures (or doped compounds) this methodology also helps to understand how further modification can be initiated for the sake of improvement in thermoelectric properties. Besides, In addition to lowering thermal conductivity, it is important to raise power factors to decent values (at least a few $\mu W.K^{-2}.m^{-1}$) for practical applications. Indeed the improvement of zT of ZnO is one of the objectives of this work, the main objective of this work is to predict the addition of the optimum amount of foreign materials (dopants and or additives) in the oxide samples to improve the power factors. Hitherto most reported composites do not provide any rationale for the mixing of two materials and rely on an empirical approach. Therefore, it becomes challenging to predict the optimum amount to be used. Therefore, in this paper, we report the method of using the Jonker and Ioffe model to provide an experimental demonstration to predict the optimum amount of foreign material (here graphite) along with the study of their thermoelectric properties. Besides, there are reports on shape-dependent thermoelectric properties of ZnO such as nano and microparticles (spheres),[17] nanorods,[46] nanoflower,[47] etc. Whereas, the thermoelectric properties of triangular ZnO microcrystals have not been studied to the best of our knowledge. Therefore, to signify the same the triangular shape has been emphasized.

## 2. Experimental details:
### 2.1. Synthesis Process:

$Zn_{1-x}Al_xO$ (x= 0, 0.1, 0.2) micro-crystal powders have been prepared using the solution combustion method.[48] 8.0 g of $Zn(NO_3)_2.6H_2O$, and 2.68 g of urea are mixed well in 10 ml of deionized water. 0, 0.102, and 0.206 g of $Al(NO_3)_3.9H_2O$ are added to achieve Al doping of x = 0, 0.1, 0.2 respectively. Further, each of these solutions is placed in a combustion furnace preheated at 500°C. Because of the exothermic combustion reaction, the mixture is burnt with a flame, and the desired powders are obtained as ash left behind. Each powder is divided into three equal amounts and then it is mixed with 0-2 wt.% of graphite in the step of 1%. Similarly, another powder is prepared by mixing 0.5 wt.% of graphite with the same batch of bare ZnO powder. The powders are pelletized into cylindrical pellets by the hydraulic uniaxial hot pressing technique to suppress the grain growth during sintering. This still achieves the densification at 1150°C under 5000 psi pressure in an argon atmosphere for 5 minutes using a graphite die having boron nitride sprayed all over the inner surface and on the bid to avoid extra graphite inclusions from the die. All the samples are highly dense. The density is around 85%-90% of the theoretical density and an image showing the cross-section of the sample is shown in supplementary information Figure. S1.

## 2.2. Sample Characterization and Measurement:

The as-prepared pellets have been examined using a Philips Panalytical X-pert Pro x-ray diffractometer having Cu $K_\alpha$ x-ray source of 0.15418 nm in the range of 25- 85 degrees and step size of 0.017 degrees. Further, the Raman spectra have been recorded using a 532 nm laser source on the Horiba Jobin Vyon LabRam HR Raman system at room temperature. High-resolution scanning electron microscopy (SEM) has been undertaken on Nova Nano SEM with an Electron Dispersive X-ray spectroscopy (EDS) analyzer. The Resistivity and Seebeck coefficient measurements have been performed using the Ulvac-riko ZEM-3. Hall measurements have been performed on the Hall Effect Measuring System (HEMS) made by Nano-magnetics. The thermal conductivity has been calculated using the equation,

$$\kappa = D.C_p.d \tag{1}$$

where, the thermal diffusivity (D) has been measured using a laser flash apparatus (LFA 1000 Laser Flash), the specific heat ($C_p$) measurement has been done on a DSC Q2000 TA differential scanning calorimeter, and the density (d) has been determined by the Archimedes' method.

## 3. Result and discussion:

### 3.1. Structural analysis:

X-ray diffraction (XRD) is performed on the powder samples to study the influence of Al doping as well as the addition of graphite in the lattice structure of ZnO. The XRD patterns of a few samples are displayed in Figure 1(a). As the graphite was mixed physically, in addition to the reflections of ZnO,[17] a graphite peak[49] is expected to appear. All the samples in the study show all the symmetry-allowed reflections and the same are marked in Figure 1(a). The inclusion of graphite shows a peak around the 2θ value of around 27°. The patterns of all the samples in the study correspond to the hexagonal wurtzite structure of ZnO. Due to the addition of Al, there is a small shift found in the peak (002) at the higher angle of 2θ and the samples' crystallinity is enhanced exhibiting sharp peaks as shown in Figure. S2 in the supplementary material section. As the radius of $Al^{3+}$ ion (0.53Å) is smaller than that of $Zn^{2+}$ ion (0.74Å), the shift is observed at the higher angle depicting reduced lattice spacing via substitution of Al at the Zn site. No significant change is observed in the position of the peak (101) for the doped ZnO compared to that of undoped samples. However, doping Al in the lattice is seen to cause the splitting in the peaks which could mark the presence of the secondary phase of $ZnAl_2O_4$ and marked with (*)[50] in Figure 1(b). In the inset of Figure 1(a), the $ZnAl_2O_4$ precipitate peaks are also marked by the arrows.[24] The bare ZnO shows a single peak at the 2θ value 36.3° for (101) reflection, whereas the same shows a shoulder as the Al content increases (see Figure 1(b)). The doping-induced reduction in spacing would result in strain in the lattice due to ionic radii mismatch, which is estimated by the well-known Williamson-Hall plot[51] by the eq. (2),

$$\beta_\tau cos\theta = \frac{\gamma\lambda}{D} + 4\varepsilon\ sin\theta \qquad \_\_\_\_\_(2)$$

where, $\beta_\tau$ = Full Width at Half Maxima (FWHM) of the peak in radians, $\gamma$ = shape factor (0.9), $\lambda$ = wavelength of the copper $K_\alpha$ source, $\theta$ = half of the peak position (Bragg angle) in radians. The relative strain in the lattice ($\varepsilon$) and the crystallite size (D) have been calculated from the slope and the intercept respectively as portrayed in Figure 1(c). The values have been tabulated in Table 1. As the crystallite size refers to the dimension of each crystal present inside a particle or grain, the measured size from SEM micrographs and calculated size from the W-H plot differ for the larger particles.

The XRD patterns of graphite and 1% Al co-doped ZnO samples are shown in Figure. S3(a) in the supplementary material section. There is a prominent change found in the peak positions in Al-doped ZnO concerning the undoped one. As an effect of graphite inclusion, the extra peak of graphite is found similar to the ZnO with graphite samples. Due to the graphite inclusions, the peaks become sharper than the only 1% Al-doped ZnO sample as shown in Figure. S3(b). In Figure. S3(c) the FWHM of three main peaks of the same samples are tabulated and the reduction in FWHM with an increase in graphite concentration confirms the improvement in crystallinity of the samples.

**Table 1. The expected and quantified formula from SEM, the strain, and the crystallite size were calculated from the W-H plot of the undoped and Al-doped ZnO samples.**

| *Sample name* | *Expected Formula* | *Quantified formula from SEM EDAX* | *Calculated Strain (%)* | *Crystallite size by W-H plot (nm)* |
|---|---|---|---|---|
| **Bare ZnO** | ZnO | $ZnO_{0.903}$ | 0.18% | $30 \pm 5$ |
| **1% Al-doped ZnO** | $Al_{0.01}Zn_{0.99}O$ | $Al_{0.013}Zn_{0.987}O_{0.863}$ | 0.22% | $28 \pm 5$ |
| **2% Al-doped ZnO** | $Al_{0.02}Zn_{0.98}O$ | $Al_{0.024}Zn_{0.976}O_{0.813}$ | 0.23% | $25 \pm 5$ |

## 3.2. Morphological Studies:

The SEM micrographs and the EDS microanalysis are done on the ZnO as well as the Al-doped ZnO samples. All the samples are seen to mainly consist of triangular crystallites of around 700-1000 nm as shown in Figure 2(a) and the magnified image of the same material is shown in Figure 2(b). Besides those triangular crystallites, a few hexagonal or semi-hexagonal crystallites, and broken crystallites of the size range of a few hundred nm are randomly distributed in the materials. By semi-hexagonal shape, it means a hexagon with 3 sides of the same length *a*, and the other 3 sides are of length *b* as shown in the above picture. It could also be called a truncated equilateral triangle as shown in the inset in Figure 2(c). In Figure 2(c), the ZnO with the 0.5 wt % of graphite sample morphology is shown. The dotted boundaries mark the triangular crystallites. The presence of layered graphite is denoted by the dotted circular boundaries.

In Figure 2(d) the EDS spectrum of 1% Al-doped ZnO is displayed confirming the presence of Aluminum in the sample. From the EDS spectrum of the without graphite samples the quantification of the Aluminum has been determined and tabulated in the inset histogram. From the measured value the nominal atomic formula has been calculated and the same is shown in Table 1. From Table 1 it may be noted that the observed composition is in close agreement with the nominal one. Nevertheless, all the samples are found to be excess in zinc than rather oxygen-deficient which is usually expected.[52] The samples were broken from the middle and SEM as well as EDS is performed on the interior side of the 1% Al-doped ZnO having 2 wt% Gr composition. The SEM micrographs are shown in Figure. S4(a) and (b). The SEM-EDS is performed at the different places of the 1AZO2G sample as shown in Figure. S4(a) and (b). It is performed mainly to confirm the existence of carbon. The weight% of the composition is shown in Figure S4(c) which reflects the presence of carbon and is almost uniform throughout the sample. This confirms the homogenous nature of the samples with the existence of graphite.

### 3.3. Raman spectroscopy

To understand the effect of graphite inclusion on the vibrational modes of bare ZnO, Raman spectroscopy has been performed on the four densified pellet samples. In Figure 3(a) and (b) the Raman spectra are presented in full range and low energy region respectively. The two distinct regions are identified: the first one is from wavenumber 200 cm$^{-1}$ to 1000 cm$^{-1}$ for the vibrational modes of ZnO and the other is 1000 cm$^{-1}$ to 3000 cm$^{-1}$ for the region of the graphite. The ZnO peaks are observed to be significantly broadened due to the smaller crystallite size. Most of the transverse optical modes (TO) are not visible in the spectra; nevertheless, longitudinal optical (LO) phonon modes are very prominent. The TO and LO modes are denoted by the red and black arrows respectively in Figure 3(b). The $B_2$ mode is observed around 275 cm$^{-1}$ is visible in the bare ZnO sample but upon addition of graphite in the sample leads to suppression of $B_2$ mode.

There are weak signals for $A_1$ (TO)[53,54] and $E_1$ (TO) mode around 385 cm$^{-1}$ and 400 cm$^{-1}$ respectively in all the ZnO-graphite samples. The peaks prominently appear in the bare ZnO sample around 520 cm$^{-1}$ and 570 cm$^{-1}$ due to the surface phonon modes (SPM). Whereas the $E_1$ (LO) mode is suppressed due to the inclusion of graphite. The peaks seen at around 1330 cm$^{-1}$ and 1580 cm$^{-1}$ indicate the presence of graphite (D and G) in the graphite-mixed samples. As the graphite content in the ZnO increases, the D and G peaks [55,56] in the Raman spectra become predominant. Ideally, in the ZnO samples, any peak for graphite is not expected; however, during hot pressing in graphite dies, some graphite may get incorporated on the surface.[45] Therefore, a very weak signal of graphite is observed in the bare ZnO as well.

### 3.4. Thermoelectric properties and statistical prediction of realizable maximum TE power factor:

The temperature-dependent electrical conductivity of all the doped/undoped and with or without Graphite compositions are shown in Figure 4(a). It is seen that for all the samples the electrical conductivity is initially almost independent of temperature and at very high temperatures, it gradually falls like a metal. However, the conductivity of the undoped ZnO without graphite composition is nearly independent of temperature for the whole temperature range. Overall the electrical conductivity is found to increase with Al doping. However, no definite trend is observed with Graphite content. The conductivity is plotted as a function of the inverse of temperature (T) in a semi-log graph. The Arrhenius plot is drawn in Figure. S5 in the supplementary material section. From the slope of the graphs, activation energies are calculated and tabulated in Table S1.

The activation energy remains the same over the entire temperature range only for the bare ZnO as the conductivity shows a nearly temperature-independent behavior in the range of temperature studied. Compared to the undoped ZnO, the one with Al doping as well as graphite inclusion, shows a sudden change in slope of conductivity vs temperature plot at around 625 K. Here, it should be noted that the Al-doped ZnO has a higher carrier concentration than that of undoped ZnO (as seen in Figure 4(c)).

This is reflected in the drop in mobility as conductivity can be described by the relation[17],

$$\sigma = ne\mu \qquad \ldots\ldots\ldots\ldots (3)$$

Where $\sigma$ is the electrical conductivity, $n$ is the carrier concentration, $e$ is the charge of the electron and $\mu$ is the mobility of the material.

From the eq. (3) it is evident that the electrical conductivity is a combined effect of the carrier concentration and the mobility. Doping of Al in the Zn site enhances the carrier concentration heavily as shown in Figure 4(b). 1% Al doping is found to enhance the carrier concentration by one order of magnitude, whereas increasing the Al content to 2%, the carrier concentration increases only by a small amount. (both 1 and 2% Al show ~$10^{20}$ /cm³). Similarly, the mobility reduces along with the increase in carrier concentration for the 1% Al-doped samples. However, the population of the charge carriers in the 2% Al-doped samples hinders the mobility heavily and reduces the conductivity ($\sigma_{300}$: conductivity at 300K). Therefore, for a degenerately doped semiconductor, there could be a significant contribution of electron-phonon scattering at higher temperatures resulting in reduced carrier mobilities (and ria se in effective masses) in Al-doped samples. The same is also seen in the case of a sudden rise in Seebeck as well as a fall in thermal conductivity measurements discussed later.

The effect of graphite inclusion is quite interesting. From the temperature-dependence of conductivity data of the ZnO-graphite mixture, it may be inferred that the addition of even 0.5 wt.% graphite increases the conductivity of bare ZnO. On increasing the graphite content to 1%, the conductivity is affected more adversely than the ZnO0.5G. On further increasing the graphite content to 2 wt%, the electrical conductivity is observed to increase and is slightly higher than the bare ZnO. An attempt is made to understand the probable mechanism and it is schematized in Figure 5. In the case of bare ZnO, the grain boundary resistance[57] is the key reason for the lower conductivity as shown in Figure 5(a).[58] When a very small amount (0.5 wt%) of the graphite is introduced, it appears between the grains of bare ZnO. This creates the downward band bending (as the work function of n-type ZnO (5.3 eV)[59] is greater than Graphite work function (4.62 eV)[60]) at the grain boundaries as shown in Figure 5(a). for ZnO0.5G. As a result, it lowers the grain boundary resistance and allows the relatively high energy charge carriers which were unable to cross the grain boundary earlier, now can move freely and block the low energy charge carriers to enhance the mobility simultaneously.[61] Therefore, the conductivity increases. However, when the graphite amount is doubled i.e. 1 wt%, it forms a wider well at the grain boundaries as shown in the case of ZnO1G. Due to the widening of the well the mid-energy carriers which already cross a grain boundary start becoming localized (less mobile) and a few of those are trapped in the well. As a combined result of these, the conductivity is reduced. In the addition of 2 wt% of graphite, the percolation paths through the graphite channels open up. In Figure. 5b, the extracted conductivity data of graphite is compared with bare ZnO, 1AZO, and 2AZO. The conductivity of graphite is fairly comparable to the conductivity of doped and bare ZnO samples. The formed graphite channels in the ZnO2G create parallel paths to conduct along with the ZnO-graphite channels that enhance the conductivity of ZnO slightly. From this discussion, it is evident that the addition of 2wt% of graphite opens up the parallel graphite channels. In Figure 5(b). it is portrayed that the conductivity of 1AZO and 2AZO without graphite is higher than the graphite conductivity.[62] Therefore, the addition of 2 wt% of graphite in the 1AZO and 2AZO lowers the conductivity than their 1wt% graphite added counterparts.

In a typical semiconductor, the conductivity falls with a drop in temperature exponentially whereas in metal conductivity increases linearly with a drop in temperature. In a composite that is made up of a semiconductor (ZnO) and metal (Graphite), the total conductivity of the composite depends on the relative conductivity difference, their phase fraction, and how they are present in the composite. For instance: there are two important aspects of conductivity i.e. direct conductivity and effective conductivity. The direct conductivity refers to the conductivity of each individual component or phase within the composite material. In a composite, different phases (e.g., different materials or grains) may have different intrinsic conductivities. Direct conductivity is typically measured or calculated for each phase separately in a homogenous phase, without considering the interactions or the overall structure of the composite. On the other hands, effective conductivity refers to the overall conductivity of the composite material as a whole that may not be just the weighted average of the constituent conductivity because of the presence of interfaces. It takes into account the distribution, orientation, and interaction of the different phases or materials within the composite. This parameter is crucial for understanding how the composite as a whole behaves in practical applications. There could be ohmic or non-ohmic interfaces (with energy barriers) at the interfaces of the constituent controlling the overall conductance. Moreover, the thermal activation to individual constituent and their interfacial charge transfer rates would have different temperature dependence. Therefore, the temperature response of electrical conductivity of Al: ZnO/graphite composites is non-trivial and complex to decipher.

In the case of undoped ZnO, there exist many defect states in the vicinity of the Fermi level due to the Zn interstitials, oxygen vacancies, etc. The Al dopants further add more donor levels. However, for the bare ZnO without graphite sample, the electrical conductivity is found to be almost independent of temperature like an extrinsic semiconductor in a saturation region where all donors or acceptors are ionized. This could be due to such lack of any more donors (defects) to be ionized above room temperature and thus there is an insignificant change in the ratio of carrier concentration and mobility with the temperature rise. However, the graphite addition to the ZnO may ease the band bending at the ZnO-ZnO interface[2] resulting in high effective donor (carrier) concentration and hence the metallic conduction. As the temperature increases further, more free carriers are generated and the fermi level moves near the conduction band. Due to this the conduction band becomes populated and as a result, the mobility decreases drastically after a certain temperature (around 600 K). For the doped samples, the carrier concentration is already high at the conduction band at room temperature. Therefore, the increase in the temperature gives rise to the carrier concentration and after the mentioned temperature the mobility reduces heavily by showing a more drastic reduction in the conductivity of doped samples. In a similar analogy, it may be explained for the graphite mixed Al-doped ZnO samples. Nevertheless, in addition to these carrier-carrier scatterings, there could be an Umklapp process-generated phonons at a high temperature which could contribute to a sudden fall in conductivity. Therefore, it requires further investigations to understand the exact transport process.

The conductivity mechanism in the samples with the addition of 1% graphite may be explained with the help of Figure 5(c). Due to the addition of Al, the carrier concentration increases which raises the Fermi level near the conduction band. The work function of the sample is reduced. As the addition of the graphite bends the band at the grain boundary, the bending reduces with the increase in the Al doping. Therefore, with the increase of doping concentration in the 1 wt% graphite mixed samples the number of free carriers increases. The detailed calculation regarding the band bending with doping concentration is given in the supplementary information and portrayed in Figure. S6 in the supplementary material section.

From the single parabolic band model, the Seebeck coefficient of metals and degenerate semiconductors can be explained by the Mott formula[17] that involves carrier concentration (n) and carrier mobility(μ). So both these quantities would govern the Seebeck coefficient of the material. The electrical conductivity and Seebeck coefficient show an opposite dependency on the concentration thus when one rises the other one falls. In this case, the measure S values as a function of temperature follow the same as shown in Figure 4(b). The negative sign of Seebeck in all the samples depicts n-type behavior in all the cases. As mentioned, the difference of an order of magnitude in the carrier concentration is reflected in the Seebeck coefficient of the undoped (> -200μV/K) ZnO reduced to (<100 μV/K) in Al-doped samples.

The expression given in equation (5) is deduced by *Ioffe* and Jonker's formulation, to estimate the highest possible power factor values of this series of materials. Once the conductivity and Seebeck values have been measured for all nine samples, a *Jonker* analysis has been performed. The procedure and the formulation are described below.[40]

The expression for S can also be written as:

$$S = -\frac{k_b}{e}\left\{ln\left(\frac{N_c}{n}\right) + A\right\} \quad \ldots\ldots\ldots\ldots (5),$$

where $k_b/e$ has a fixed value of 86.15 μV/K, and $N_c$ is the conduction band Density of states (DOS).

$\left(N_c = \left(\frac{2\pi^3 m^* k_b T}{h^2}\right)^{3/2}\right)$, and A is the transport constant (0 ≤ A ≤ 4)[40]. From eq. (3) and eq. (5) The relation between Seebeck and electrical conductivity could be easily derived. i.e.

$$S = \frac{k_b}{e} \ln\left(\frac{\sigma}{\sigma_0}\right) = \frac{k_b}{e}\{\ln\sigma - \ln\sigma_0\} \qquad \text{............(6)}$$

where $\sigma_0 = N_v.e.\mu.\exp(A)$ and $N_v$ is the DOS of the valence band for the n-type material. The negative sign in the Seebeck coefficient confirms the n-type nature of the series of this material. Now the power factor (PF) can be calculated from the eq. (7).

$$PF = S^2\sigma \qquad \text{................. (7)}$$

This leads to the result:

$$PF = \frac{k_b^2}{e^2}\left[\ln\left(\frac{\sigma}{\sigma_0}\right)\right]^2 . \sigma \qquad \text{............... (8),}$$

and, from the *Ioffe* formula, the optimization can be done as,

$$\frac{\partial(PF)}{\partial(\ln\sigma)} = \left(\frac{k_b}{e}\right)^2 \exp(\ln\sigma)(\ln\sigma - \ln\sigma_0)(\ln\sigma - \ln\sigma_0 + 2) \qquad \text{.............. (9)}$$

Now, to estimate the maximum PF the eq. (9) should equal to 0. It leads to the result $\ln\sigma = \ln\sigma_0$ or $\ln\sigma = \ln\sigma_0 - 2$. The first one gives the PF as 0 which is unphysical, so the second one is the physically relevant solution. Therefore,

$$PF_{max} = 4\left(\frac{k_b}{e}\right)^2 (\ln\sigma_0 - 2) \qquad \text{..................... (10)}$$

$$\ln PF_{max} = \ln\sigma_0 + \ln 4\left(\frac{k_b}{e}\right)^2 - 2 \approx -19.33 + \ln\sigma_0 \qquad \text{............... (11)}$$

To determine the $PF_{max}$ first a plot of S vs $\ln\sigma$ is drawn from the measured data. A straight line of slope 86.15 µV/K is fitted in the undeviated region as shown in Figure 4(d) and $\ln\sigma_0$ is extracted from the intercept of the $\ln\sigma$ axis. This value is now put in the eq. (11) to estimate the maximum possible value of the power factor i.e. around 450 µW/m-K². The *Jonker* plot also gives an idea about the degeneracy of the material. As from Figure 4(d), it is also obtained that when the lnσ value becomes more than 10.5 for these materials the deviation occurs because the S value starts saturating irrespective of the conductivity value. It is denoted by the arrow in the same figure. Later on, the thermal conductivity ($\kappa$) of these materials is measured, and based on that the *figure of merit,* zT was calculated by the well-known formula of the thermoelectric figure of merit i.e.

$$zT = \frac{S^2 \sigma}{k} \qquad \qquad \ldots\ldots\ldots\ldots (12)$$

Based on the measurement there is not much change due to the graphite inclusion. Bare ZnO displays the best zT among these materials. It has a power factor of around 260 µV/mK². The inclusion of graphite reduced the overall efficiency when the same or above 1 wt. % is introduced in the material. To achieve the predicted power factor value bare ZnO is modified by adding 0.5 wt. % of graphite and the measurement is pursued. This inclusion doubled the PF value and achieved the desired PF of the material as shown in Figure 4(d). Later on, this was found to be a higher thermally conducting material than the bare ZnO however, the overall efficiency improved.

The thermal conductivity consists of two parts i.e. the electronic part ($\kappa_e$) and the lattice part ($\kappa_l$). The electronic part can be obtained using the S and the σ values given in eq. (14) and (15).

$$\kappa = \kappa_e + \kappa_l \qquad \qquad \ldots\ldots\ldots\ldots (13)$$

$$\kappa_e = L\sigma T \qquad \qquad \ldots\ldots\ldots\ldots (14)$$

Here, L is the Lorenz number. It may be calculated using the eq. (15).

$$L = 1.5 + exp\left(-\frac{|S|}{116}\right) \times 10^{-8} \qquad \qquad \ldots\ldots\ldots\ldots (15)$$

The $\kappa_l$ can be determined from eq. (13). The total, lattice, and electronic thermal conductivity and the power factor are shown in Figure 6(a), (b), (c), and (d) respectively.

The electronic part of the thermal conductivity follows the trend of the electrical conductivity. It showed a downfall after around 625 K as expected. Figure S7 in supplementary Material shows the comparison of the thermal conductivity of the samples near room temperature (335 K). It may be seen that the total thermal conductivity is dominated by the lattice contribution in all the samples irrespective of doping or addition. On average, compared to pristine ZnO, ZnO with graphite additives exhibits higher lattice thermal conductivity. This could be due to the addition of a secondary phase (graphite) which itself has a very high thermal conductivity[63]. The bare ZnO has high lattice thermal conductivity compared to its doped counterparts. However, the incorporation of even a small amount of graphite (0.5 wt%) is seen to be present uniformly throughout the ZnO matrix. The process of physically mixing the ZnO and Graphite in a mortar pestle should lead to the exfoliation of graphite and hence create thin layers of graphite between two ZnO grains. Such a few-layer graphene could have a high thermal conduction contribution. Therefore, here the ZnO-graphite interface plays a crucial role and the entire composite may be modeled as a network of thermal resistors of two kids, one is ZnO and the other is graphite. The net thermal conduction is established due to the least resistive or cumulative thermal resistance of the network. There are few reports on the arbitrary behavior of lattice thermal conductivity upon graphite addition into materials like oxides,[32] chalcogenides,[64] and intermetallics[33], etc. These results demonstrate that the thermal conductivity of composites with graphite has a statistical variation in thermal conductivity dependent on microstructure evolution from sample to sample in interface-controlled composites. As the temperature increases, the lattice part of the thermal conductivity is reduced due to several scattering mechanisms. However, the rate of decrement in the lattice thermal conductivity increased after around 575 K except for the bare ZnO sample. If only the scattering process dominates, this should have followed a similar trend but as these materials are a mixture of two different materials the phonons might have changed their group velocity ($V_g$) when they propagated through a different chemical environment (from ZnO grain to graphite grain) or some secondary point defects (Al doping in ZnO). This phenomenon is known as lattice softening, shown in the inset of Figure 6(b). It mainly dominates at temperatures when T > $\Theta_D$ (Debye Temperature) where phonon-phonon scattering is the dominating scattering process. The $\Theta_D$ for ZnO is around 390 K. Above this temperature the scattering process and lattice softening dominated simultaneously but after around 575 K the lattice softening process dominates and it results in the faster change in the slope of the lattice thermal conductivity. Only the bare ZnO is without external defects and after the formation of the composite with graphite, it follows the trend of the Umklapp scattering process.

The lattice thermal conductivity ($\kappa_l$) of a material above its Debye temperature can be expressed by the Slack model[65] as:

$$\kappa_l = \frac{A v_g^3}{T} \qquad \ldots\ldots\ldots (16)$$

therefore,

$$\frac{d\kappa_l}{dT} = -\frac{A v_g^3}{T^2} + 3A \frac{v_g^2}{T}\left(\frac{dv_g}{dT}\right) \qquad \ldots\ldots\ldots (17)$$

where A is the constant depending upon the material parameters, $v_g$ is the phonon group velocity and T is the temperature.

The possible uneven dip in the thermal conductivity at higher temperatures can be due to the second term in equation (xvii) involving the rate of change in phonon group velocity with temperature. The lattice softening can occur due to lattice defects such as interstitials, vacancies, dislocations, and the presence of a foreign material. These defects introduce the change in phonon frequency locally within the material that leads to the lattice softening. Besides, phonon-phonon scattering, grain boundary scattering, or electron-phonon scattering the change in sound velocity due to the presence of this lattice softening also scatters phonons and assists in reducing the thermal conductivity. Specifically, at high temperatures for anharmonic materials the lattice softening is expected to dominate and the lattice thermal conductivity deviates (reduces) from the expected values. Therefore, in the ZnO sample the presence of interstitial defects, in doped samples the presence of Al doping as a defect along with the impure phase of $ZnAl_2O_4$ as foreign material and in graphite mixed samples the presence of graphite as foreign material along with other defects lead to the lattice softening effect and reduce the lattice thermal conductivity mainly at high-temperature region. An effort has been made to fit the measured data with the Slack model to show the mentioned change for the undoped ZnO samples mixed with the graphite of different concentrations. The fitted and the measured data are shown in Figure S8 in the supporting information.

Finally, based on the measured electronic and thermal properties the zT is calculated using eq. (12) and plotted in Figure 7. The bare ZnO with the addition of 0.5 wt% of graphite shows the best zT of 0.093 (near zT = 0.1) among all the prepared samples.

The Hot press method does not allow the powder to loosen to its minimum energy supplied during the pellet formation. Along with the uniaxial pressure it helps to achieve the desired densification without any notable grain growth. Due to the use of the graphite die in this densification process some amount of graphite came to the surface of the sample. Due to this factor, the D and G Raman peaks were observed in the bare ZnO sample slightly. On the other hand, the ZnO samples with graphite are with the above peaks with high intensity. Even a slight rise of the 2D peak is also observed in the samples as a signal of the formation of graphene. It helped to improve the surface conductivity. However, the effect of graphite inclusion does not help to increase the electrical conductivity monotonously because of the dual effect of enhancement in the carrier density and the mobility of the charge carriers and also the formation of the secondary phase of $ZnAl_2O_4$. In this series of materials, doping has intensified the carrier density but mobility decreases more than expected. So as a combined effect, it does not serve the purpose of enhancing efficiency. A very small amount of graphite helps to increase the efficiency of the ZnO. The plot based on the Jonker and Ioffe formulation helps to achieve the most possible power factor. The introduction of secondary material in the mixture shows the effect of lattice softening when the carriers diffuse through it. As a result, the thermal conductivity has reduced at a higher temperature than the desired reduction. As a result, it helped to magnify the zT.

Hence keeping in mind all of these parameters the ZnO with 0.5 wt% graphite sample displays higher zT value as the effect of carrier density enhancement and less mobility reduction. The inclusion also helps to reduce the thermal conductivity by phonon scattering and as well as lattice softening.

## 4. Conclusion

Undoped, 1, and 2% Al-doped ZnO microparticles are prepared by the combustion method. Then parts of the synthesized powder are mixed without and with 1 wt% and 2 wt% graphite. From the electrical and thermal measurements, the values of the power factor are calculated. Based on the Seebeck and electrical conductivity values the *Jonker* plot is drawn to predict the highest achievable power factor along with the use of *Ioffe* formulation. The graphical calculation delivers a predicted power factor value of 450 μV/K. Later, analyzing the power factor and thermal conductivity data of the prepared samples a new sample of ZnO mixed with 0.5 wt% graphite is made and measured. It reaches near the value of the predicted highest power factor value of 425 µV/K and achieves the highest zT of 0.093 at 900 K as well. From the point of view of efficiency, this is an average group of materials. However, it is challenging to optimize the zT when there is more than one parameter to optimize. Starting with a few samples as the trial samples and performing the thermoelectric measurements and analysis the prediction of the highest power factor can be done with the help of *Jonker* and *Ioffe* plots. This method is applied to these samples and the predicted value is achieved. Changing the doping concentration along with the amount of graphite inclusion becomes an arbitrary method to improve the zT. Moreover, with the help of the above methods, the expected results are experimentally achieved for this set of samples.

## ASSOCIATED CONTENT

**Supplementary Information**

Further analysis of XRD data, Microstructure of the Al-doped ZnO with Graphite inclusions, Band diagram of ZnO-graphite, Comparison of thermal conductivity and its analysis.

**Corresponding Author**

Vinayak Kamble kbvinayak@iisertvm.ac.in

**Author Contributions**

The manuscript was written through contributions of all authors. All authors have given approval to the final version of the manuscript.

**Funding Sources**

ACKNOWLEDGMENT

The authors are thankful to the thermal conductivity measurement facility, IITB, and Prof. Satish Vitta for his support. The Hall effect measurement facility CIF IISERTVM is gratefully acknowledged.

**Data availability statement**

The data is available with the corresponding author upon a reasonable request.

**Conflict of interest**

The authors declare no conflict of interest.

**ORCID IDs**

Soumya Biswas 0000-0003-3587-8596

Keshav Dabral 0000-0001-7499-4093

Saptak Majumder 0000-0002-2140-1140

Rajasekar Parasuraman 0000-0002-9240-7574

Aditya S Dutt 00000-0003-0663-8905

Vinayak Kamble 0000-0002-6344-2194

List of Figures

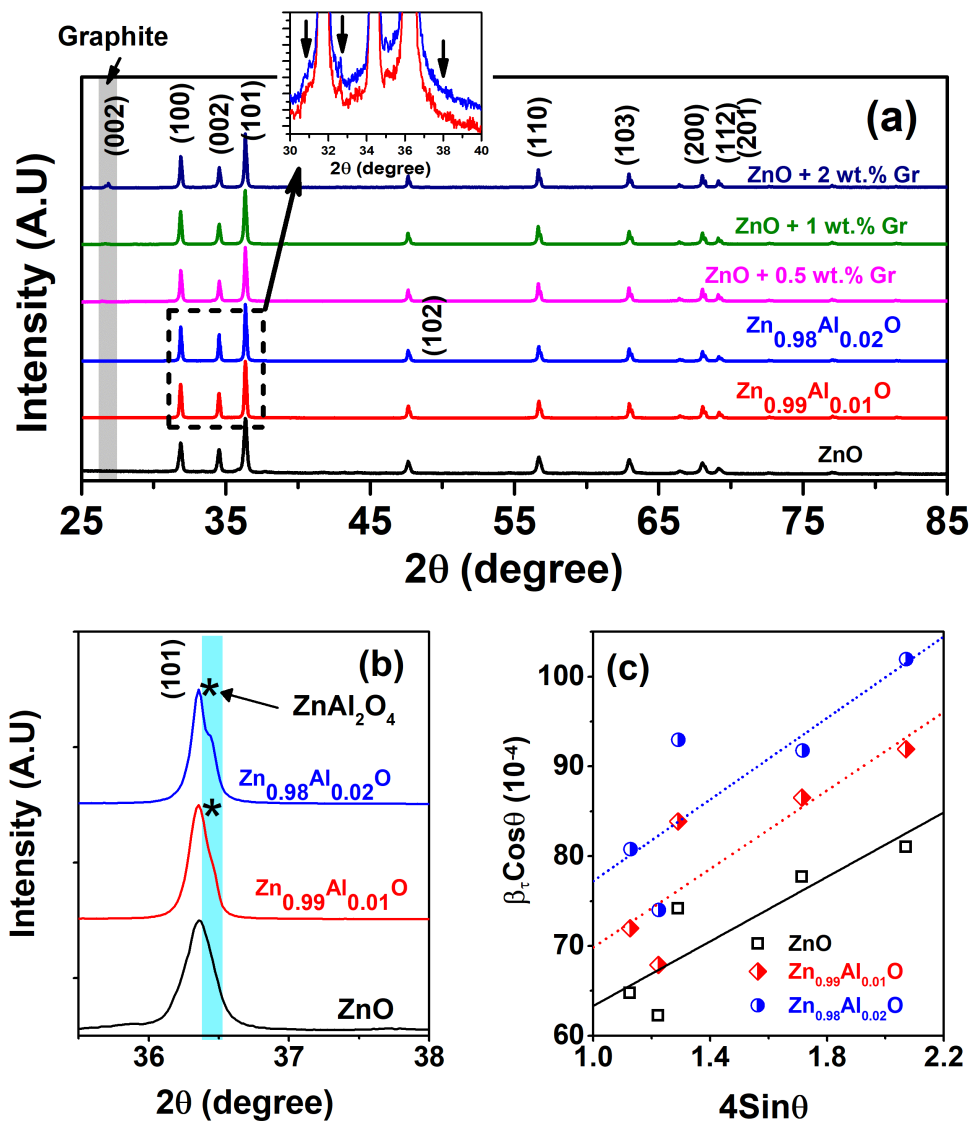

**Figure 1. (a) The XRD pattern of the undoped, Al-doped ZnO samples and their composites with graphite, (the inset shows the ZnAl$_2$O$_4$ precipitate peaks) (b) The enlarged view of ZnAl$_2$O$_4$ shoulder to ZnO peak seen in the XRD of Al-doped (without graphite) samples. (c) The Williamson-Hall plot of the doped and undoped samples to calculate the strain present in the samples.**

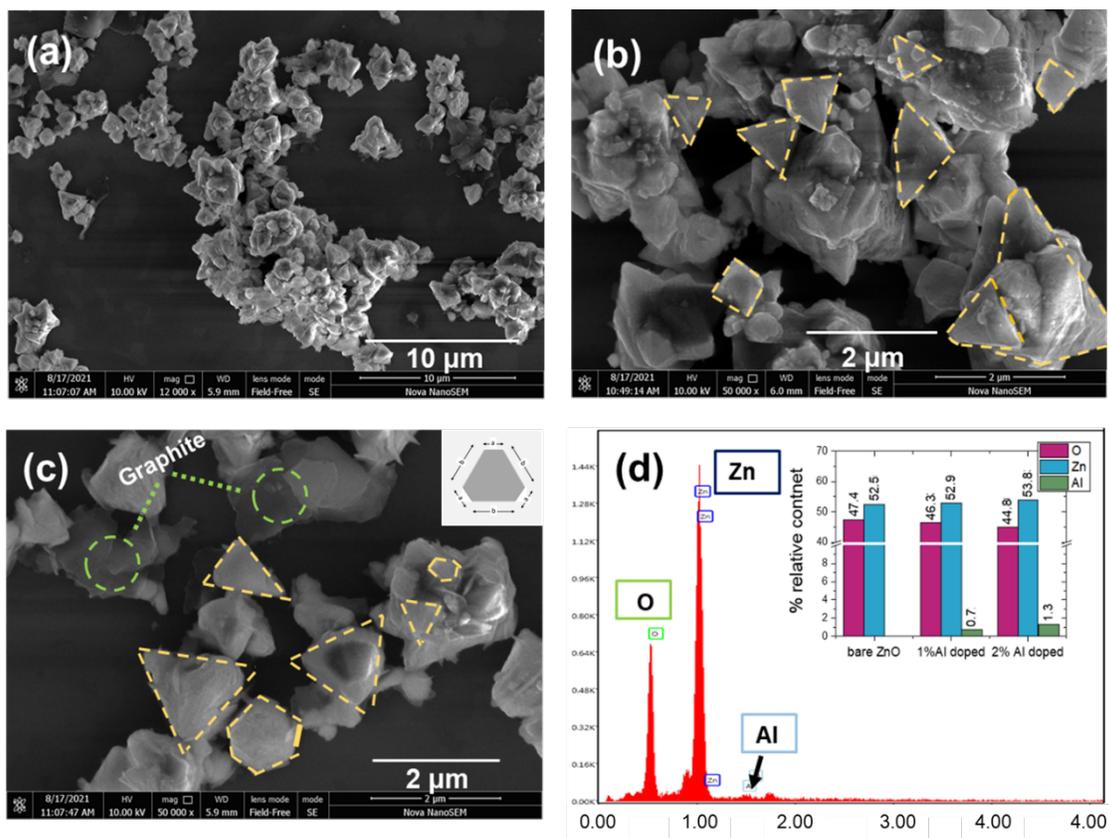

**Figure 2.** The SEM micrographs of (a) bare ZnO, (b) bare ZnO in higher magnification with prominent triangular microcrystals denoted by yellow borders, (c) ZnO0.5G with microcrystals and graphite denoted by green circles. A semi-hexagonal shape is shown inset (d) the EDS spectrum of 1AZO. In inset shows the quantification of the elements in the undoped and doped ZnO samples.

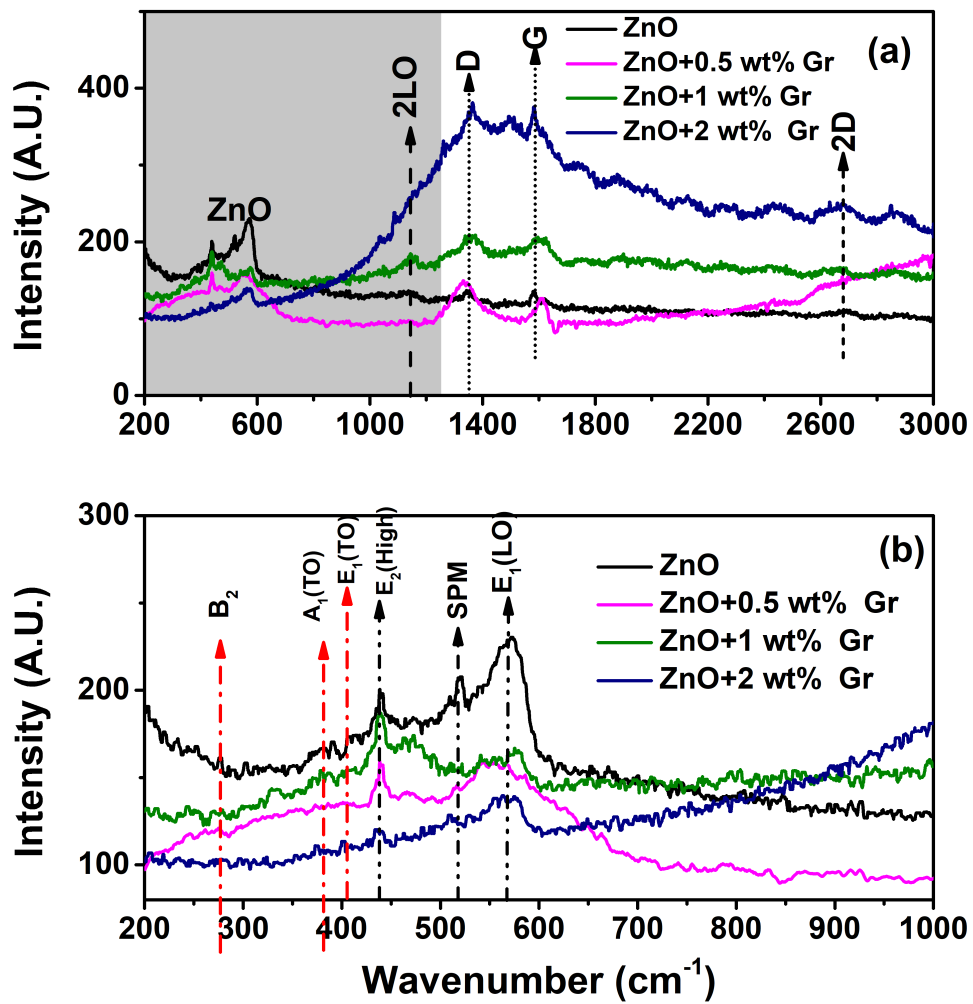

**Figure 3.** The Raman shift in the range of (a) 200-3000 cm$^{-1}$ and (b) 200-1000 cm$^{-1}$ (to magnify ZnO peaks) for the undoped ZnO samples with different graphite concentrations.

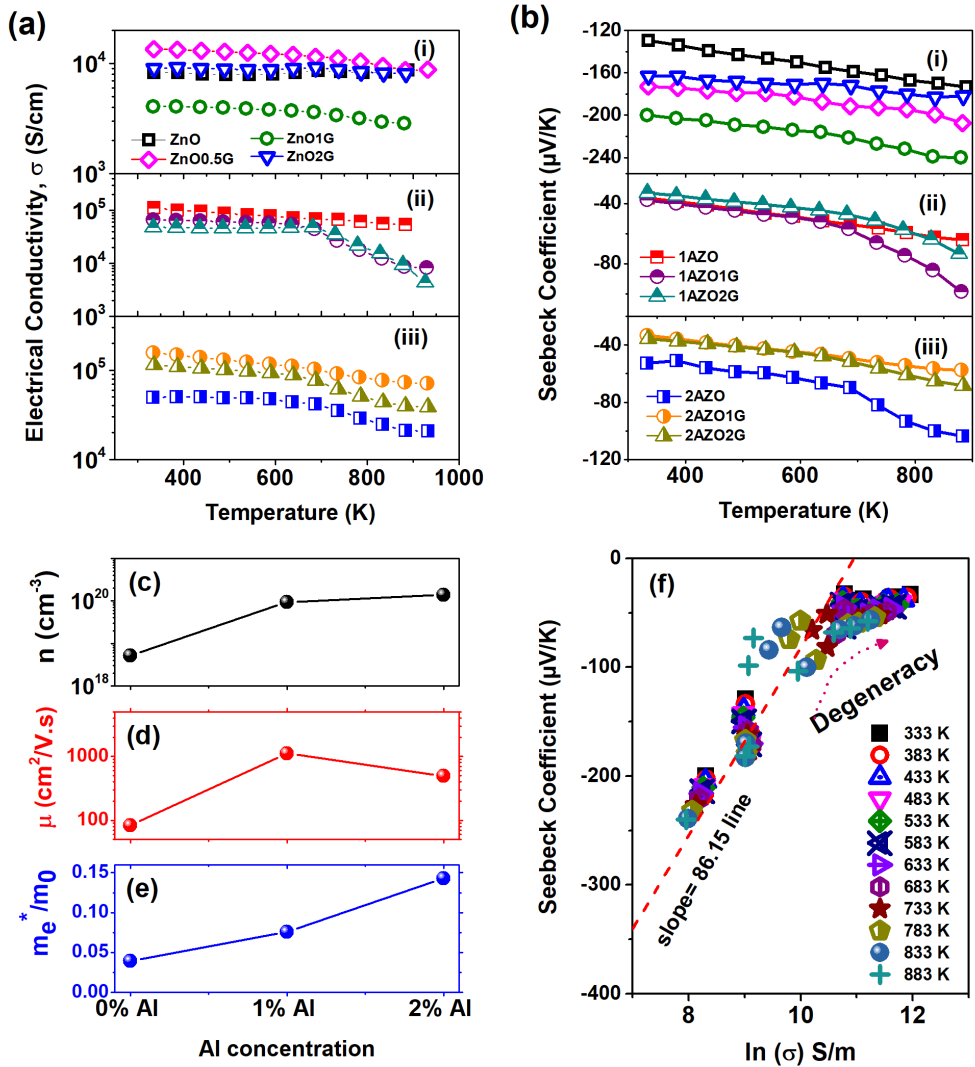

**Figure 4.** (a) Electrical conductivity, σ and (b) Seebeck coefficient plotted as a function of temperature for all the (i) undoped, (ii) 1% Al doped and (iii) 2% Al doped ZnO samples mixed with graphite. The measured (c) carrier concentration, (d) mobility and (e) effective mass variation with Al concentration at 300 K ($\sigma_{300}$). (f) the Jonker plot for all the undoped, doped, graphite mixed ZnO samples.

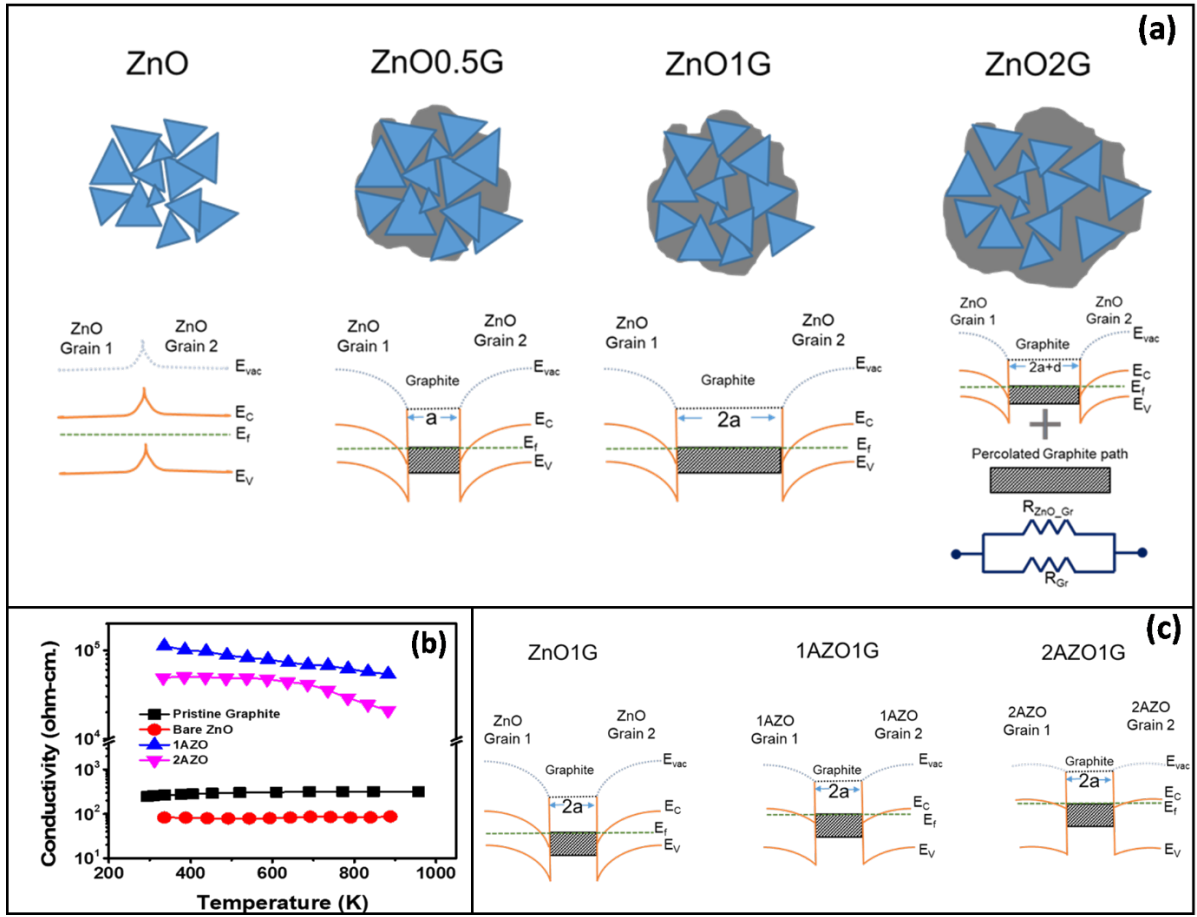

**Figure 5. (a) Schematic diagram of conduction mechanism in ZnO with graphite samples (b) the comparison of the conductivity of bare ZnO, 1AZO, 2AZO, and Graphite (c) Schematic diagram of band bending in the same amount of graphite mixed samples with different doping concentrations of Al.**

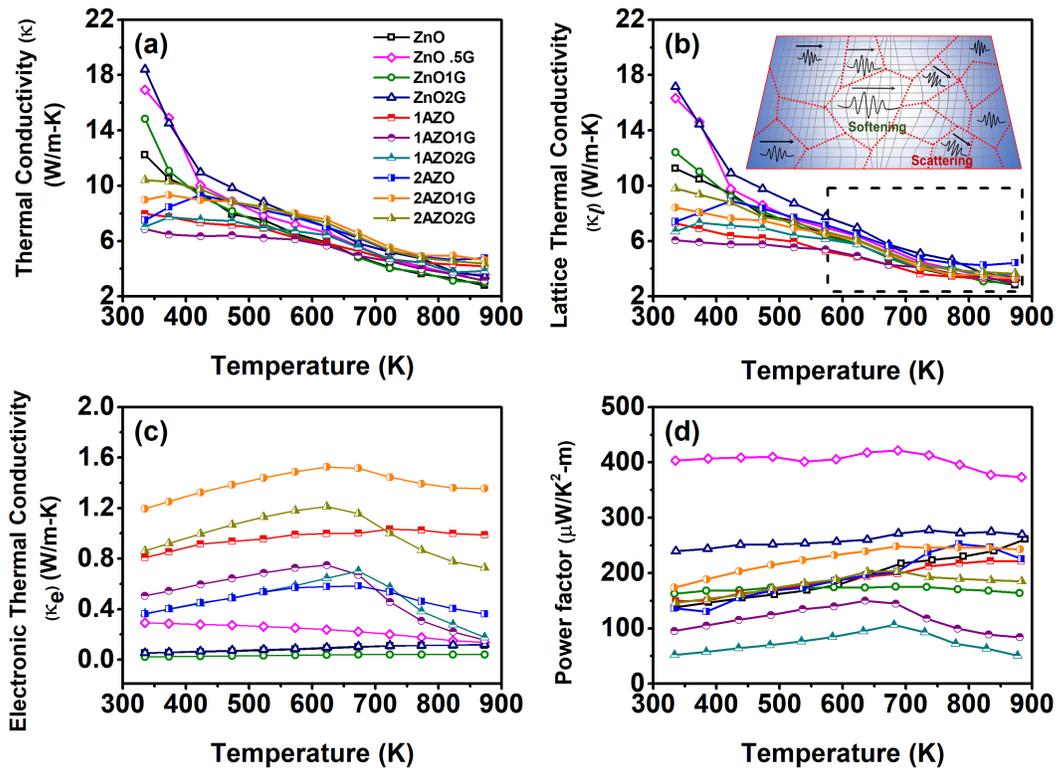

**Figure 6.** The (a) total, (b) lattice (c) electronic thermal conductivity, and (d) power factor of all ten samples. In the inset of (b) the schematic of the lattice softening.

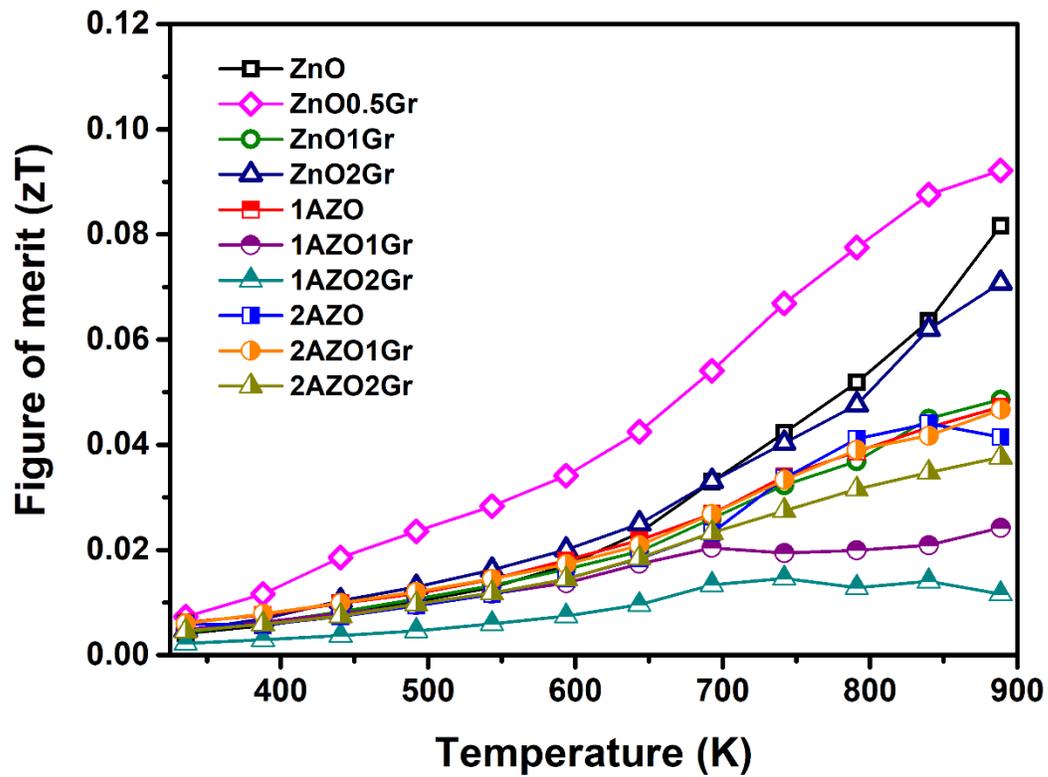

**Figure 7.** The Figure of merit (zT) of all the prepared samples ZnO graphite composite samples.